\documentclass[a4paper,prl,showpacs,preprint]{revtex4}

\usepackage{times}
\usepackage{amssymb}
\usepackage{graphicx}

\begin{document}

\title{Influence of low energy scattering on loosely bound states}
\author{Jean-Marc Sparenberg}
\email{jmspar@ulb.ac.be}
\author{Pierre Capel}
\author{Daniel Baye}
\affiliation{Physique Quantique,
Physique Nucl\'eaire Th\'eorique et Physique Math\'ematique,
Universit\'e Libre de Bruxelles (ULB), CP 229, B 1050 Bruxelles, Belgium}
\date{\today}

\begin{abstract}
Compact algebraic equations are derived,
which connect the binding energy and the asymptotic normalization constant (ANC) of a subthreshold bound state with the effective-range expansion of the corresponding partial wave.
These relations are established for positively-charged and neutral particles,
using the analytic continuation of the scattering (S) matrix in the complex wave-number plane.
Their accuracy is checked on simple local potential models for the $^{16}$O+n, $^{16}$O+p and $^{12}$C+$\alpha$ nuclear systems,
with exotic nuclei and nuclear astrophysics applications in mind.
\end{abstract}


\maketitle

There is a renewal of interest today for the quantum description of the low-energy scattering of two particles.
This interest is mostly triggered by condensed-matter and ultracold-gases physics
but similar studies are performed in other fields.
Nuclear physicists in particular are studying low-energy collisions in the context of nuclear astrophysics and of exotic nuclei.
A problem specific to nuclear physics is that positively-charged particles
repel each other,
which makes cross sections extremely small at low energy,
and hence hard or impossible to measure.
Theory thus plays an important role here but is
made more complicated by the Coulomb interaction.

Particular problems occur in the presence of weakly bound states:
because of the wave nature of quantum phenomena, these play a role similar to resonances, leading to huge variations of cross sections.
Such subthreshold bound states occur both in neutral cases
(e.g., magnetic Feshbach resonances in atom-atom collisions or the ``historical'' deuteron bound state)
and in charged cases.
Famous examples are the lowest $2^+$ and $1^-$ states of the $^{16}$O nucleus which,
lying just below the $^{12}$C+$\alpha$ threshold,
are thought to strongly influence the $^{12}$C($\alpha$,$\gamma$)$^{16}$O capture cross section and hence the carbon to oxygen ratio in red giant stars \cite{azuma:94,brune:99}.
Essential quantities required for the theoretical description of this reaction are the binding energy of these states and their ANC,
which characterizes the tail of their slowly-decreasing wave function
[see Eq.\ (\ref{C}) below].
While the energy can be precisely measured,
the ANC is still rather poorly known:
it is not directly accessible experimentally
and various indirect methods have been proposed to infer it.
Among them, a high-precision measurement of elastic scattering is believed to give crucial information \cite{tischhauser:02} but the ANC extraction from this measurement relies on a reaction-matrix analysis \cite{angulo:00}, which is made rather delicate by the description of the non-resonant cross section \cite{sparenberg:04a}.

In the present work, a more fundamental approach is proposed to relate scattering properties to bound-state properties, in particular to the ANC, in the case of a weakly bound state.
This approach is based on general S-matrix properties \cite{mukhamedzhanov:99}:
non-resonant scattering states are described with the help of the effective-range expansion while bound states are described in terms of poles of the S matrix in the complex plane.
Combining these allows us to derive compact algebraic equations which prove particularly useful and promising, as illustrated by the examples below.
In the following, we mostly concentrate on the charged case
but also discuss the simpler neutral case, which leads to interesting comparisons.

Let us consider two particles of charges $Z_1$, $Z_2$ and of reduced mass $\mu$.
We denote the center-of-mass energy by $E=\hbar^2 k^2/2\mu$
and the dimensionless Sommerfeld parameter by $\eta=1/a_B k$,
where $k$ is the relative wave number
and $a_B=\hbar^2/Z_1 Z_2 e^2 \mu$ is the nuclear Bohr radius.
The wave number is allowed to be complex, the upper-half complex plane corresponding to the physical energy sheet \cite{joachain:83}.
The scattering matrix $S_l$ for a partial wave $l$ in the presence of both a Coulomb and a short range (e.g., nuclear) interaction is defined by \cite{newton:82}
\begin{equation}
 S_l(k)=e^{2i \sigma_l} e^{2 i \delta_l}
 = \frac{\Gamma(l+1+i \eta)}{\Gamma(l+1-i \eta)} \times
\frac{\cot \delta_l(k) + i}{\cot \delta_l(k) - i}.
 \label{Sl}
\end{equation}
In these expressions, the first factor is the pure Coulomb scattering matrix,
with $\sigma_l(k)=\arg \Gamma(l+1+i \eta)$ being the Coulomb phase shift.
The second factor is due to the additional short-range interaction,
with $\delta_l$ being the additional phase shift.

In the following, we are interested in low physical energies (either positive or negative) and hence would like to expand the functions of interest around the origin $E=0$.
Since the S matrix defined above has a rather complicated analytical structure in the complex energy plane, due to the Coulomb interaction, we follow Refs.\ \cite{hamilton:73,vanhaeringen:77} and introduce a function with simpler analyticity properties,
\begin{eqnarray}
 F_l(k^2) & = & \frac{e^{2i \delta_l(k)}-1}{2i} \times \frac{l!^2 e^{2i\sigma_l(k)} e^{\pi \eta}}
 {k^{2l+1}\Gamma^2(l+1+i \eta)} 
 \label{FlS} \\
 & = & \frac{1}{\cot \delta_l(k) -i} \times \frac{l!^2 a_B^{2l+1} (e^{2\pi\eta}-1)}{2\pi w_l(\eta^2)}, 
 \label{Fldel}
\end{eqnarray}
where two alternative expressions are given for both factors.
In the last equation,
we have used
\begin{equation}
 w_l(\eta^2)=\prod_{n=0}^{l}\left(1+\frac{n^2}{\eta^2}\right).
\label{wl}
\end{equation}
Next, we define on the physical energy sheet \cite{hamilton:73,iwinski:84}
\begin{equation}
 h(\eta^2)=\psi(i \eta)-\ln(i\eta)+ (2i\eta)^{-1},
 \label{h}
\end{equation}
where $\psi$ is the digamma function \cite{abramowitz:65}.
This allows constructing the {\em effective-range function}
\begin{equation}
 K_l(k^2)=\frac{1}{F_l(k^2)}+\frac{2 w_l(\eta^2)}{l!^2 a_B^{2l+1}} h(\eta^2),
 \label{Kl} 
\end{equation}
which is holomorphic in the physical energy sheet, with a cut along part of the negative real axis \cite{hamilton:73}.
This function is analytic and regular at the origin, which implies it can be Taylor expanded [see Eq.\ (\ref{Klser}) below].
Moreover, it is real for real energies.

In the neutral case, things are much simpler:
by taking $\eta=0$ in Eqs.\ (\ref{Sl}) and (\ref{FlS}),
the Coulomb factors simplify and the additional phase shift reduces to the usual phase shift. An effective-range function,
with the same properties as in the charged case,
can be defined as \cite{joachain:83}
\begin{equation}
K_l(k^2)=\frac{1}{F_l(k^2)}+i k^{2l+1} \quad (\eta=0).
 \label{Kl_neutral} 
\end{equation}

Let us now connect the effective-range function with physical properties for both positive and negative energies.
For positive energies, it is related to the scattering phase shift.
In the charged case,
Eq.\ (\ref{Fldel}) yields \cite{hamilton:73,kamouni:07}
\begin{equation}
 K_l(k^2)=\frac{2 w_l(\eta^2)}{l!^2 a_B^{2l+1}}\left[\frac{\pi}{e^{2\pi\eta}-1}\cot \delta_l(k)+\Re h(\eta^2)\right],
 \label{ere}
\end{equation}
where the imaginary part has vanished because of the value of the digamma function for an imaginary argument \cite{abramowitz:65}.
In the neutral case, Eq.\ (\ref{FlS}) implies \cite{joachain:83}
\begin{equation}
 K_l(k^2)=k^{2l+1} \cot \delta_l(k) \quad (\eta=0).
 \label{ere_neutral}
\end{equation}

For negative energies, bound states correspond to poles of the S matrix on the positive imaginary $k$ axis \cite{joachain:83}.
Let $k=i \kappa_b$ be the location of such a pole, $E_b=-\hbar^2\kappa_b^2/2\mu$ and $\eta_b=-i/a_B \kappa_b$ be the corresponding binding energy and Sommerfeld parameter.
Equation (\ref{FlS}) shows that this bound state corresponds to a zero of $F_l^{-1}$ and Eq.\ (\ref{Kl}) implies that \cite{iwinski:84}
\begin{equation}
 K_l(-\kappa_b^2) = \frac{2  w_l(\eta_b^2)}{l!^2 a_B^{2l+1}} h(\eta_b^2).
\label{ereE}
\end{equation}
In the neutral case, Eq.\ (\ref{Kl_neutral}) implies instead that
\begin{equation}
 K_l(-\kappa_b^2) = (-1)^{l+1} \kappa_b^{2l+1} \quad (\eta=0).
\label{ereE_neutral}
\end{equation}
These equations relate the effective-range function to the bound-state energy.
Though Eq.\ (\ref{ereE_neutral}) is frequently used in the literature,
in particular to treat the $l=0$ deuteron bound state \cite{joachain:83},
Eq.\ (\ref{ereE}) is less known.

The ANC $C_b$ of this bound state is defined
by the asymptotic behavior of its normalized radial wave function $R_b(r)$, i.e.,
\begin{equation}
 R_b(r)\mathop{\sim}_{r\rightarrow \infty} C_b \exp(-\kappa_b r)/r^{|\eta_b|+1}.
\label{C}
\end{equation}
Under specific conditions,
this ANC is related to the residue of the S-matrix bound-state pole by \cite{baz:69,joachain:83,mukhamedzhanov:99}
\begin{equation}
 S_l(k) \mathop{\sim}_{k\rightarrow i \kappa_b} (-1)^{l+1} i e^{-\pi \eta_b}
 \frac{|C_b|^2}{k-i \kappa_b}.
 \label{SC}
\end{equation}
In Ref.\ \cite{blokhintsev:08},
this relation is shown to be valid for potentials decreasing faster than $\exp(-2\kappa_b r)$.
It thus holds, e.g., for all bound states of cut-off potentials,
and for weakly bound states of potentials decreasing fast enough,
a condition assumed in the present work.
In other cases, in particular for deeply bound states of exponentially decreasing potentials,
Eq.\ (\ref{SC}) can be violated \cite{sparenberg:04a}.
Combining Eqs.\ (\ref{Sl}), (\ref{FlS}), and (\ref{SC}) implies that \cite{iwinski:84}
\begin{equation}
 |C_b|= \kappa_b^l \frac{\Gamma(l+1+|\eta_b|)}{l!} \left[\left.-\frac{d F_l^{-1}}{dk^2}\right|_{k^2=-\kappa_b^2}\right]^{-\frac{1}{2}},
 \label{ereC}
\end{equation}
which provides a second relationship between the effective-range function and the bound-state properties.


Equations (\ref{ereE}), (\ref{ereE_neutral}), and (\ref{ereC}) are quite general;
let us now particularize them to small binding and scattering energies by using the Taylor expansions of the various holomorphic functions introduced above for $k^2 \rightarrow 0$ (i.e., for $\eta^2 \rightarrow \infty$).
The expansion of function $w_l$ directly ensues from its definition (\ref{wl});
function $h$ expands as
\begin{equation}
 h(\eta^2) = \frac{1}{12 \eta^2} + \frac{1}{120 \eta^4} + \frac{1}{252 \eta^6} + O\left(\frac{1}{\eta^8}\right)
\end{equation}
and we write the first terms of function $K_l$ as
\begin{equation}
 K_l(k^2) = -\frac{1}{a_l}+\frac{r_l}{2} k^2 - P_l r_l^3 k^4 + Q_l k^6 + O(k^8).
 \label{Klser}
\end{equation}

The pole location conditions (\ref{ereE}) and (\ref{ereE_neutral}) lead to different relations for the charged and neutral cases.
Truncating Eq.\ (\ref{ereE}) to the first order in $E$ leads to
\begin{equation}
 \frac{1}{a_l} \approx \left(-\frac{r_l}{2} + \frac{1}{6 a_B^{2l-1} l!^2}\right) \kappa_b^2,
 \label{ere1E}
\end{equation}
while truncating Eq.\ (\ref{ereE_neutral}) leads to
\begin{equation}
  \frac{1}{a_l} \approx -\frac{r_l}{2} \kappa_b^2 + (-1)^l \kappa_b^{2l+1} \quad (\eta=0).
 \label{ere1E_neutral}
\end{equation}
Both equations directly relate the scattering length $a_l$ to the binding energy,
provided the effective range $r_l$ is known.
They imply that the scattering length is large when the binding energy is small,
$r_l$ being in general rather independent of the binding energy.
Note that, for $l=0$, Eq.\ (\ref{ere1E_neutral}) is the well-known Schwinger relation \cite{bethe:49};
for a very small binding energy, the second term then dominates, which implies that the scattering length becomes independent of the effective range.
For $l>0$, on the contrary, the first term dominates and Eq.\ (\ref{ere1E_neutral}) becomes closer to Eq.\ (\ref{ere1E}).

Combining these equations with Eq.\ (\ref{ereC}) provides
useful expressions for the ANC in terms of the scattering length.
For the charged case, one gets
\begin{equation}
 |C_b|  \frac{l!}{\Gamma(l+1+|\eta_b|)} \approx \kappa_b^{l+1} \sqrt{a_l}
\propto \kappa_b^l,
 \label{ere1C}
\end{equation}
which generalizes Eq.\ (45) of Ref.\ \cite{mukhamedzhanov:99} to $l \ge 0$.
This equation is actually also valid for the neutral case with $l>0$,
where the fraction of the first member simplifies.
For $l=0$, one gets the very different result \cite{joachain:83}
\begin{equation}
 |C_b| \approx \sqrt{2/a_0} \approx \sqrt{2 \kappa_b} \quad (\eta=l=0).
 \label{ere1C_00}
\end{equation}

\begin{figure} [t]
\rotatebox{-90}{\scalebox{0.44}{\includegraphics{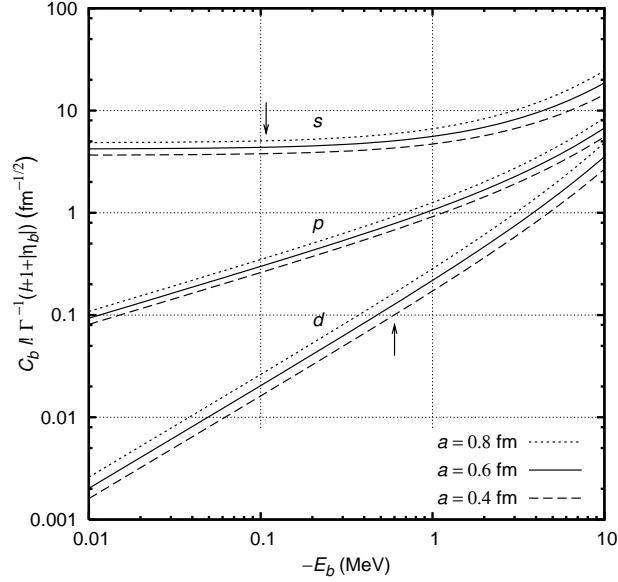}}}
 \caption{\label{CE} ANC of a schematic $^{16}$O+p bound state as a function of its binding energy, for the $l=0$, 1 and 2 partial waves and three different values of the potential diffuseness $a$. The two $^{17}$F physical bound states are indicated by arrows.}
\end{figure}

These simple equations clearly illustrate the possibility of directly extracting the ANC from the binding energy and the elastic-scattering phase shifts.
However, let us stress that their validity is limited to low energies,
since they rely on a first-order effective-range expansion in energy.
Higher-order expansions can be obtained [see, e.g., Eq.\ (\ref{ere3C}) below]
but lead to rather heavy expressions not reproduced here for brevity.
Above a few terms, a purely numerical treatment is probably most appropriate.
Let us also note that the precision on the ANC expected from these equations
strongly depends on the precision available on the scattering length.
In the presence of a subthreshold bound state,
we have seen that the scattering length is very large.
Since it appears in the effective-range expansion (\ref{Klser}) with a negative power,
it has a much smaller influence on this expansion than the following terms.
Hence, it is typically difficult to determine $a_l$ from scattering phase shifts in such a case.
Equations like (\ref{ere1E}) and (\ref{ere1E_neutral})
then come as a useful tool:
they allow one to calculate $a_l$ from the binding energy and the next terms of the expansion,
which are easier to determine from scattering experiments.

\begin{figure} [t]
\rotatebox{-90}{\scalebox{0.44}{\includegraphics{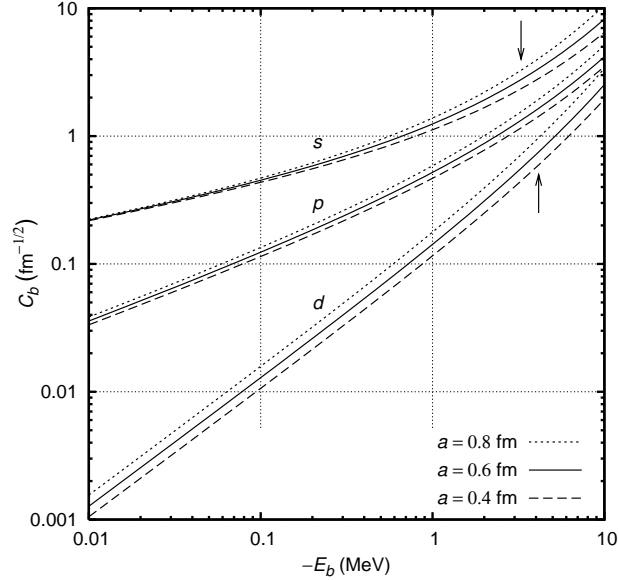}}}
 \caption{\label{CE_neutral} Same as Fig.\ \ref{CE}, but for the $^{16}$O+n system.}
\end{figure}

Let us now check the accuracy of these results on typical nuclear physics examples.
We first study the binding-energy dependence of ANCs for Woods-Saxon potentials of radius 3.023 fm, with and without Coulomb interactions.
These potentials schematically represent the $^{16}$O+p and $^{16}$O+n systems.
For several values of the potential diffuseness,
we vary the depth of these potentials,
in order to scan a wide range of binding energies.
For each binding energy, we calculate numerically the ANC.
The results for the $s$, $p$ and $d$ waves are presented in Figs.\ \ref{CE} for the charged case and \ref{CE_neutral} for the neutral case.
The noticeable simplicity of these curves up to about 1 MeV can be easily explained by Eqs.\ (\ref{ere1E})--(\ref{ere1C_00}):
the ANCs have the expected energy dependence,
showing the small sensitivity of the effective range $r_l$
to the potential depth.
On the contrary, its sensitivity to the potential diffuseness is larger and increases with the magnitude of the barrier due to the Coulomb and/or centrifugal potentials.
For the remarkable $s$-wave neutral case [Eq.\ (\ref{ere1C_00})],
the behavior at small binding energy becomes independent of the potential,
a situation also encountered for the deuteron bound state \cite{bethe:49}.

The $^{17}$F and $^{17}$O nuclei both have an $s$ and a $d$ bound state,
well represented by the above potentials.
Their binding energies are indicated by arrows in Figs.\ \ref{CE} and \ref{CE_neutral}.
For the $^{17}$F nucleus, the $s$ state is weakly bound,
which suggests a one-proton halo structure.
Quantitatively, Eq.\ (\ref{ere1C}) provides the ANC of this state with an accuracy better than 0.1\%.
Equation (\ref{ere1E}) predicts $a_0$ with an error of 5\%,
while a generalization to the second order in energy reduces this error to 0.4\%,
which shows the fast convergence of our method.
Now, for this system, a second-order effective-range expansion turns out to be sufficient to fit experimental scattering data on a rather wide energy range (a few MeV) \cite{kamouni:07}.
The present results thus show that both the scattering length and the bound-state ANC could be precisely deduced from these data.


Let us now turn to a more important example from the physical point of view.
The first $2^+$ excited state of $^{16}$O, which lies 245 keV below the $^{12}$C+$\alpha$ threshold, plays a key role in the calculation of the $^{12}$C$(\alpha,\gamma)^{16}$O radiative capture cross section \cite{azuma:94,brune:99}.
This state is known to have a $d$-wave $^{12}$C+$\alpha$ structure and is reasonably well described by a $^{12}$C+$\alpha$ potential model \cite{sparenberg:04a}.
Here, we choose a nuclear Gaussian potential inspired by Ref.\ \cite{buck:84},
which reads $-112.3319 \exp(-r^2/2.8^2)$ MeV,
where $r$ is the internucleus distance in fm,
and a Coulomb potential $12 e^2 \mathrm{erf} (r/2.5)/r$,
where erf is the error function \cite{abramowitz:65}.
This potential has a bound state at $E_b=-245.0$ keV, corresponding to $\kappa_b=0.1876$ fm$^{-1}$ with $\hbar^2/2\mu=6.964$ MeV fm$^2$ and $a_B=0.8060$ fm.
Numerically, one finds the ANC $C_b = 1.384 \times 10^5$ fm$^{-1/2}$.
Let us now try to recover this value with Eq.\ (\ref{ere1C}).
The numerical method of Ref.\ \cite{baye:00b}
provides $a_2=5.891 \times 10^4$ fm$^5$, $r_2=0.1580$ fm$^{-3}$, and $P_2=-65.96$ fm$^8$
for this potential.
Equation (\ref{ere1C}) then leads to $C_b \approx 1.23 \times 10^5$ fm$^{-1/2}$ (11\% error),
while Eq.\ (\ref{ere1E}) provides $a_2 \approx 4.76 \times 10^4$ fm$^5$ (19\% error).
A generalization of these equations to higher orders
thus seems necessary in this case.
Up to the third order in energy, the ANC formula reads
\begin{equation}
 |C_b| \approx \frac{\kappa_b^3 \Gamma(3+|\eta_b|)/2}{\sqrt{\frac{1}{a_2}-\left(P_2 r_2^3+\frac{17}{80 a_B}\right)\kappa_b^4-2\left(Q_2-\frac{191 a_B}{1008}\right)\kappa_b^6}}.
\label{ere3C}
\end{equation}
By neglecting the last two terms of the denominator, one recovers Eq.\ (\ref{ere1C}).
Adding the second term leads to $C_b \approx 1.414 \times 10^5$ fm$^{-1/2}$ (3\% error).
To determine the third term, we need the next coefficient of the effective-range expansion;
for that, we have fitted the numerical phase shifts with Eqs.\ (\ref{ere}) and (\ref{Klser}),
which provides $Q_2=0.1411(2)$ fm.
Adding this term strikingly improves the fit:
within a 1\% precision, the phase shifts are fitted up to 80 keV without it,
and up to 700 keV with it.
This improvement is very promising as experimental data,
which are only available above 750 keV \cite{plaga:87},
could be fitted with a limited number of terms in the effective-range expansion.
With the three terms, Eq.\ (\ref{ere3C}) provides $C_b \approx 1.379 \times 10^5$ fm$^{-1/2}$ (0.4\% error),
a remarkable precision for a third-order effective-range expansion.

In conclusion, starting from general S-matrix properties,
we have established equations connecting the effective-range function
with the energy and ANC of bound states,
for an arbitrary partial wave and for both charged and neutral systems.
By using the effective-range series expansion,
we have deduced from these relations compact equations [Eqs.\ (\ref{ere1E})--(\ref{ere3C}), possibly extended to higher orders]
that relate the energy and ANC of a subthreshold bound-state to the series coefficients
(scattering length, effective range\dots).
We have checked on test cases that these relations are precise and could be used
to extract a subthreshold bound state ANC from scattering data in a model-independent way
when precise experimental phase shifts are available.
We plan to apply our method to the $^{12}$C+$\alpha$ experimental phase shifts of Ref.\ \cite{tischhauser:02} in a near future.

\acknowledgments{
PC is supported by the F.R.S.-FNRS.
This text presents research results of the Belgian Research Initiative on eXotic nuclei (BriX), program P6/23 on interuniversity attraction poles of the Belgian Federal Science Policy Office.
}


\begin{thebibliography}{19}
\expandafter\ifx\csname natexlab\endcsname\relax\def\natexlab#1{#1}\fi
\expandafter\ifx\csname bibnamefont\endcsname\relax
  \def\bibnamefont#1{#1}\fi
\expandafter\ifx\csname bibfnamefont\endcsname\relax
  \def\bibfnamefont#1{#1}\fi
\expandafter\ifx\csname citenamefont\endcsname\relax
  \def\citenamefont#1{#1}\fi
\expandafter\ifx\csname url\endcsname\relax
  \def\url#1{\texttt{#1}}\fi
\expandafter\ifx\csname urlprefix\endcsname\relax\def\urlprefix{URL }\fi
\providecommand{\bibinfo}[2]{#2}
\providecommand{\eprint}[2][]{\url{#2}}

\bibitem[{\citenamefont{Azuma et~al.}(1994)\citenamefont{Azuma, Buchmann,
  Barker, Barnes, D'Auria, Dombsky, Giesen, Jackson, King, Korteling
  et~al.}}]{azuma:94}
\bibinfo{author}{\bibfnamefont{R.~E.} \bibnamefont{Azuma}}
  \bibnamefont{{\em et~al.}}, \bibinfo{journal}{Phys.\ Rev.\ C}
  \textbf{\bibinfo{volume}{50}}, \bibinfo{pages}{1194} (\bibinfo{year}{1994}).

\bibitem[{\citenamefont{Brune et~al.}(1999)\citenamefont{Brune, Geist,
  Kavanagh, and Veal}}]{brune:99}
\bibinfo{author}{\bibfnamefont{C.~R.} \bibnamefont{Brune}}
  \bibnamefont{{\em et~al.}},
  \bibinfo{journal}{Phys.\ Rev.\ Lett.} \textbf{\bibinfo{volume}{83}},
  \bibinfo{pages}{4025} (\bibinfo{year}{1999}).

\bibitem[{\citenamefont{Tischhauser et~al.}(2002)\citenamefont{Tischhauser,
  Azuma, Buchmann, Detwiler, Giesen, G{\"o}rres, Heil, Hinnefeld, K{\"a}ppeler,
  Kolata et~al.}}]{tischhauser:02}
\bibinfo{author}{\bibfnamefont{P.}~\bibnamefont{Tischhauser}}
  \bibnamefont{{\em et~al.}}, \bibinfo{journal}{Phys.\ Rev.\ Lett.}
  \textbf{\bibinfo{volume}{88}}, \bibinfo{pages}{072501}
   (\bibinfo{year}{2002}); {\em ibid.},
%
  \textbf{\bibinfo{volume}{79}}, \bibinfo{pages}{055803}
  (\bibinfo{year}{2009}).

\bibitem[{\citenamefont{Angulo and Descouvemont}(2000)}]{angulo:00}
\bibinfo{author}{\bibfnamefont{C.}~\bibnamefont{Angulo}} \bibnamefont{and}
  \bibinfo{author}{\bibfnamefont{P.}~\bibnamefont{Descouvemont}},
  \bibinfo{journal}{Phys.\ Rev.\ C} \textbf{\bibinfo{volume}{61}},
  \bibinfo{pages}{064611} (\bibinfo{year}{2000}).

\bibitem[{\citenamefont{Sparenberg}(2004)}]{sparenberg:04a}
\bibinfo{author}{\bibfnamefont{J.-M.} \bibnamefont{Sparenberg}},
  \bibinfo{journal}{Phys.\ Rev.\ C} \textbf{\bibinfo{volume}{69}},
  \bibinfo{pages}{034601} (\bibinfo{year}{2004}).

\bibitem[{\citenamefont{Mukhamedzhanov and Tribble}(1999)}]{mukhamedzhanov:99}
\bibinfo{author}{\bibfnamefont{A.~M.} \bibnamefont{Mukhamedzhanov}}
  \bibnamefont{and} \bibinfo{author}{\bibfnamefont{R.~E.}
  \bibnamefont{Tribble}}, \bibinfo{journal}{Phys.\ Rev.\ C}
  \textbf{\bibinfo{volume}{59}}, \bibinfo{pages}{3418} (\bibinfo{year}{1999}).

\bibitem[{\citenamefont{Joachain}(1983)}]{joachain:83}
\bibinfo{author}{\bibfnamefont{C.~J.} \bibnamefont{Joachain}},
  \emph{\bibinfo{title}{Quantum Collision Theory}}
  (\bibinfo{publisher}{North-Holland}, \bibinfo{address}{Amsterdam},
  \bibinfo{year}{1983}), \bibinfo{edition}{3rd} ed.

\bibitem[{\citenamefont{Newton}(1982)}]{newton:82}
\bibinfo{author}{\bibfnamefont{R.~G.} \bibnamefont{Newton}},
  \emph{\bibinfo{title}{Scattering Theory of Waves and Particles}}
  (\bibinfo{publisher}{Springer}, \bibinfo{address}{New York},
  \bibinfo{year}{1982}), \bibinfo{edition}{2nd} ed.

\bibitem[{\citenamefont{Hamilton et~al.}(1973)\citenamefont{Hamilton, Overb\"o,
  and Tromborg}}]{hamilton:73}
\bibinfo{author}{\bibfnamefont{J.}~\bibnamefont{Hamilton}},
  \bibinfo{author}{\bibfnamefont{I.}~\bibnamefont{Overb\"o}}, \bibnamefont{and}
  \bibinfo{author}{\bibfnamefont{B.}~\bibnamefont{Tromborg}},
  \bibinfo{journal}{Nucl.\ Phys.} \textbf{\bibinfo{volume}{B60}},
  \bibinfo{pages}{443} (\bibinfo{year}{1973}).

\bibitem[{\citenamefont{van Haeringen}(1977)}]{vanhaeringen:77}
\bibinfo{author}{\bibfnamefont{H.}~\bibnamefont{van Haeringen}},
  \bibinfo{journal}{J.\ Math.\ Phys.} \textbf{\bibinfo{volume}{18}},
  \bibinfo{pages}{927} (\bibinfo{year}{1977}).

\bibitem[{\citenamefont{Iwinski et~al.}(1984)\citenamefont{Iwinski, Rosenberg,
  and Spruch}}]{iwinski:84}
\bibinfo{author}{\bibfnamefont{Z.~R.} \bibnamefont{Iwinski}},
  \bibinfo{author}{\bibfnamefont{L.}~\bibnamefont{Rosenberg}},
  \bibnamefont{and} \bibinfo{author}{\bibfnamefont{L.}~\bibnamefont{Spruch}},
  \bibinfo{journal}{Phys.\ Rev.\ C} \textbf{\bibinfo{volume}{29}},
  \bibinfo{pages}{349} (\bibinfo{year}{1984}).

\bibitem[{\citenamefont{Abramowitz and Stegun}(1965)}]{abramowitz:65}
\bibinfo{editor}{\bibfnamefont{M.}~\bibnamefont{Abramowitz}} \bibnamefont{and}
  \bibinfo{editor}{\bibfnamefont{I.~A.} \bibnamefont{Stegun}}, eds.,
  \emph{\bibinfo{title}{Handbook of Mathematical Functions}}
  (\bibinfo{publisher}{Dover}, \bibinfo{address}{New York}, \bibinfo{year}{1965}).

\bibitem[{\citenamefont{Baz' et~al.}(1969)\citenamefont{Baz', Zel'dovich, and
  Perelomov}}]{baz:69}
\bibinfo{author}{\bibfnamefont{A.~I.} \bibnamefont{Baz'}},
  \bibinfo{author}{\bibfnamefont{Y.~B.} \bibnamefont{Zel'dovich}},
  \bibnamefont{and} \bibinfo{author}{\bibfnamefont{A.~M.}
  \bibnamefont{Perelomov}}, \emph{\bibinfo{title}{Scattering, Reactions and
  Decay in Nonrelativistic Quantum Mechanics}} (\bibinfo{publisher}{IPST}, \bibinfo{address}{Jerusalem}, \bibinfo{year}{1969}),
  \bibinfo{note}{translated from Russian by Z.\ Lerman}.

\bibitem[{\citenamefont{Blokhintsev and Yeremenko}(2008)}]{blokhintsev:08}
\bibinfo{author}{\bibfnamefont{L.~D.} \bibnamefont{Blokhintsev}}
  \bibnamefont{and} \bibinfo{author}{\bibfnamefont{V.~O.}
  \bibnamefont{Yeremenko}}, \bibinfo{journal}{Phys.\ At.\ Nucl.}
  \textbf{\bibinfo{volume}{71}}, \bibinfo{pages}{1219} (\bibinfo{year}{2008})
  \bibinfo{note}{[Yad.\ Fiz.\ {\bf 71}, 1247 (2008)]}.

\bibitem[{\citenamefont{Bethe}(1949)}]{bethe:49}
\bibinfo{author}{\bibfnamefont{H.~A.} \bibnamefont{Bethe}},
  \bibinfo{journal}{Phys.\ Rev.} \textbf{\bibinfo{volume}{76}},
  \bibinfo{pages}{38} (\bibinfo{year}{1949}).

\bibitem[{\citenamefont{Kamouni and Baye}(2007)}]{kamouni:07}
\bibinfo{author}{\bibfnamefont{R.}~\bibnamefont{Kamouni}} \bibnamefont{and}
  \bibinfo{author}{\bibfnamefont{D.}~\bibnamefont{Baye}},
  \bibinfo{journal}{Nucl.\ Phys.\ A} \textbf{\bibinfo{volume}{791}},
  \bibinfo{pages}{68} (\bibinfo{year}{2007}).

\bibitem[{\citenamefont{Buck and Rubio}(1984)}]{buck:84}
\bibinfo{author}{\bibfnamefont{B.}~\bibnamefont{Buck}} \bibnamefont{and}
  \bibinfo{author}{\bibfnamefont{J.~A.} \bibnamefont{Rubio}},
  \bibinfo{journal}{J.\ Phys.\ G} \textbf{\bibinfo{volume}{10}},
  \bibinfo{pages}{L209} (\bibinfo{year}{1984}).

\bibitem[{\citenamefont{Baye et~al.}(2000)\citenamefont{Baye, Hesse, and
  Kamouni}}]{baye:00b}
\bibinfo{author}{\bibfnamefont{D.}~\bibnamefont{Baye}},
  \bibinfo{author}{\bibfnamefont{M.}~\bibnamefont{Hesse}}, \bibnamefont{and}
  \bibinfo{author}{\bibfnamefont{R.}~\bibnamefont{Kamouni}},
  \bibinfo{journal}{Phys.\ Rev.\ C} \textbf{\bibinfo{volume}{63}},
  \bibinfo{pages}{014605} (\bibinfo{year}{2000}).

\bibitem[{\citenamefont{Plaga et~al.}(1987)\citenamefont{Plaga, Becker, Redder,
  Rolfs, Trautvetter, and Langanke}}]{plaga:87}
\bibinfo{author}{\bibfnamefont{R.}~\bibnamefont{Plaga}}
  \bibnamefont{{\em et~al.}},
  \bibinfo{journal}{Nucl.\ Phys.\ A} \textbf{\bibinfo{volume}{465}},
  \bibinfo{pages}{291} (\bibinfo{year}{1987}).

\end{thebibliography}
\end{document}